\documentclass[%
 reprint, twocolumn,
 amsmath,amssymb,
 aps,
]{revtex4}

\usepackage{graphicx}
\usepackage{dcolumn}
\usepackage{bm}

\usepackage{xcolor}

\begin{document}

\preprint{APS/123-QED}

\title{Signal of Cosmic Strings in Cross-Correlation of\\
21-cm Redshift and CMB Polarization Maps}
\thanks{A footnote to the article title}%

\author{Matteo Blamart}
\email{matteo.blamart@mail.mcgill.ca}
\author{Hannah Fronenberg}
 \email{hannah.fronenberg@mail.mcgill.ca}
 \author{Robert Brandenberger}
 \email{rhb@physics.mcgill.ca}
\affiliation{Physics Department, McGill University, 3600 University Str., Montreal, QC, H3A2T8, Canada
\\
{\it and} McGill Space Institute 
}%


\date{\today}

\begin{abstract}

We study the signal of cosmic string wakes present before the time of reionization in the cross-correlation signal of 21-cm redshift and B-mode CMB polarization maps. The specific non-Gaussian signal of strings in the position space cross-correlation maps can be extracted by means of a matched filtering analysis. Signals of strings with tension somewhat lower than those corresponding to the current upper bound can be identified when embedded in a background of Gaussian fluctuations from a Planck best-fit LCDM model.

\end{abstract}

\maketitle


\section{Introduction and Motivation}

According to our current understanding, matter is described (at energy densities much lower than the Planck scale) by a quantum field theory. Quantum field theories can admit topological defects solutions of various topologies: planar, linear or point-like. In the same way that different metals form defects with different dimensionality when they are cooled below the crystallization temperature, in quantum field theories defects will form after a symmetry breaking phase transition. Causality implies \cite{Kibble} that a network of defects inevitably will form during this phase transition and persist to the present time. Defects carry energy and hence lead to cosmological signatures (see e.g. \cite{VS, HK, RHBCSrev} for reviews of the cosmology of topological defects). Field theories with domain wall defects are in fact ruled out \cite{DW} since a single domain wall crossing our current cosmological horizon would overclose the universe \footnote{Assuming that the energy scale of the phase transition is higher than the scale explored in accelerator experiments.}. Similarly, theories yielding monopole defects are highly constrained \cite{monopole}. Theories with linear defects (cosmic strings) are interesting since the fractional contribution of the defect network to the total energy density of matter is independent of time.

Cosmic strings are characterized by a single number, their energy per unit length $\mu$ which we typically parametrize as $G \mu$, where $G$ is Newton's gravitational constant, and we are using natural units in which the speed of light, Planck's constant and Boltzmann's constant are set to $1$. In typical models, the value of $\mu$ is given by the square of the symmetry breaking energy scale $\eta$. Hence, searching for the signatures of cosmic strings in the sky provides a way to probe particle physics beyond the Standard Model from top dpwn (i.e. from higher energy scales downwards) as opposed to accelerator searches which probe models from bottom up. Detecting the signatures of a cosmic string in the sky would have an important impact on particle physics in that it would imply that the theory which describes matter at high energies belongs to the class which has string solutions \cite{RHBshortRev}. Currently, cosmological observations have not seen any signatures of strings, leading to an upper bound on $\mu$, and ruling out all high scale models of {\it Grand Unification} type with string solutions. Improving the upper bound would constrain larger classes of models. Detecting the signatures of a cosmic string in the sky would have important implications for astrophysics since strings may help solve a number of mysteries such as the origin of super-massive black holes at high redshifts \cite{Bryce}.

Cosmic strings cannot have any ends. Hence, they are either infinite or else closed loops. We divide the network into strings with curvature radius larger than the horizon (the ``long string'' network) and the set of loops with radius smaller than the horizon. Analytical arguments \cite{VS, HK, RHBCSrev} indicate that the network of cosmic strings at late times takes on a {\it scaling solution} according to which the statistical properties of the distribution of the strings are independent of time if all lengths are scaled to the horizon length. This is supported by numerical simulations \cite{numerical}. There is less agreement on the distribution of loops. According to some studies \cite{numerical}, the distribution of loops also scales, while according to some field theory simulations \cite{Hind} the loops rapidly decay. In this paper we will focus on the cosmological signatures of the long string network.

String contribute to the spectrum of density fluctuations. Because of the scaling distribution of the string network, the contribution to the spectrum of fluctuations is scale-invariant (see e.g. \cite{Zeld, Vil, TB} for early work). However, since the strings network evolves in time, the induced fluctuations are active and incoherent, and hence do not lead to acoustic oscillations in the CMB angular power spectrum \cite{Pen}. Based on the observed acoustic oscillations, an upper bound on the string tension can be derived \cite{CMBlimits} which is of the order
\begin{equation}
G \mu \, < \, 10^{-7} \, ,
\end{equation}
where the exact value depends on the precise number of long strings per horizon volume.

Since space perpendicular to a long string is conical \cite{cone}, strings yield a distinct lensing signal in CMB temperature anisotropy maps - lines in the sky across which the CMB temperature jumps \cite{KS}. These specific non-Gaussian features in position space can be searched for in CMB temperature maps using specific statistics such as edge detection algorithms \cite{Canny}, wavelets and curvelets \cite{Hergt} or machine learning tools \cite{Oscar}. 

Long strings moving through the intergalactic gas generate overdensities in their wake \cite{wake}, as described in the following section. These overdense regions (denoted ``wakes'') lead to distinctive position space signatures in CMB polarization maps \cite{Holder1} and 21-cm redshift maps \cite{Holder2} (see also \cite{Pagano}). Both signals are due to photons from the last scattering surface passing through a string wake on their way to us, and they are hence perfectly correlated in position space. Individually, both the polarization signal and the 21-cm signals are dwarfed by noise, but the noise contributions are uncorrelated to first approximation. The idea we explore in this paper is whether we can better extract the string signals from maps containing background ``noise'' of Gaussian fluctuations predicted in a Planck-normalized LCDM model \cite{Planck} by using cross-correlation statistics as opposed to studying the signals of the individual maps. Since the string signals are correlated in the two maps while the Gaussian signals are not (to a first approximation) we expect the string signal to be more clearly visible.

The the next section, we review cosmic string wakes. In Sections \ref{sec3} and \ref{sec4} we review the string signals in CMB polarization and 21-cm maps, respectively.  Section \ref{sec5} is the heart of the paper in which we define the cross-correlation map and introduce the statistics which we use to extract the string signal. The results of our analysis are presented in Section \ref{sec6}.

As mentioned before, we use natural units in which the speed of light, Planck's constant and Boltzmann's constant are set to $1$. We work in the context of a spatially flat homogeneous and isotropic cosmological model with scale factor $a(t)$, where $t$ is physical time. The comoving spatial coordinates are denoted by $x$, and $z(t)$ is, as usual, the cosmological redshift.

\section{Cosmic String Wakes} \label{sec2}

In this section, we outline the production mechanism of cosmic string wakes as well as their evolution. 
\subsection{Introduction about Cosmic String Wakes }

A cosmic string wake is a region of overdensity of matter created in the wake of a moving cosmic string. The formation of this wake is related to the conical structure of the space-time in the space perpendicular to the string \cite{cone} and to the transverse speed of the string. The conical structure gives a deficit angle:
\begin{equation}
\alpha \, = \, 8\pi G\mu \, ,
\end{equation}
with $\mu$ the string tension and G Newton's constant. If the string has a transverse velocity relative to the intergalactic gas, this causes an accretion of matter towards the plane behind the moving string which is determined by the velocity kick to gas particles whose magnitude is
\begin{equation}
\delta v \, = \, 4\pi \gamma_{s} v_{s} G \mu \,
\end{equation}
 with $v_{s}$ the velocity of the string (in units of the speed of light), and $\gamma_{s}$ the corresponding relativistic gamma factor.

To study the effects and signatures of wakes produced by strings, the toy model first introduced in \cite{toymodel} is used. It is based on the fact that the distribution of strings takes on a scaling solution, with the typical curvature radius of the infinite strings tracking the Hubble radius, and each Hubble volume containing of the order one of strings. Hence, the infinite string network is approximated by a set (of the order one per Hubble volume) of finite Hubble length segments which evolves between the time of recombination and today. The period of time is divided into Hubble time intervals. This distribution of segments is taken to be statistically independent at each Hubble step, i.e., at each time interval a new distribution of moving segments is randomly generated \footnote{This is justified since the infinite string network maintains its scaling form by string intercommutations and chopping off of string loops.} The speed of the segments is also randomly generated between 0 and 1 in units of the speed of light.  A cosmic  string segment laid down at time $t_{i}$ of  length $c_{1}  t_{i} $ creates a wake of dimension:
\begin{equation}
c_{1}t_{i}\times t_{i}\gamma_{s} v_{s}\times 4\pi G \mu t_{i} \gamma_{s} v_{s} \, ,
\end{equation}
where $c_{1} $is a factor of the order of 1 coming from numerical simulation of evolution of cosmic strings and is determined by the correlation length of the string network as a function of the Hubble radius.
\begin{figure}[h]
\begin{center}
\includegraphics[scale=0.46]{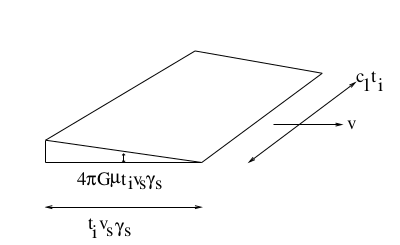} 
\caption{Sketch of cosmic string wake geometry. The wedge shown here is depicts the wake. The string producing this wake is along the thinnest part of the wedge (the line indicated by the vector $c_1 t_i$). The string is moving with a velocity $v$ indicated by the vector $v$, thus creating an overdensity of depth $t_i v_s \gamma_s$, and a thickness determined by the string tension (what is indicated in the figure is the mean thickness).}

\end{center}
\end{figure}

\subsection{Cosmic String Wake evolution}

Once a wake is formed at time $t_i$, the planar size of the wake increases as space expands. The thickness of the wake has an additional growth due to gravitional accretion. To describe this evolution, consider planes of particles with initial comoving distance $q$ from the central plane of the wake. The corresponding physical height of the plane is:
\begin{equation}
h(q,t) \, = \, a(t)[q-\psi (q,t)]
\end{equation}
with $\psi (q,t)$ the comoving displacement due to the gravitational attraction to the wake. In the Zel'dovich approximation in the Newtonian limit, the evolution of the physical height between the time $t_{i}$ and $t$ is: 
\begin{equation}
h(t,t_{i}) \, = \, \psi_{0} \dfrac{z_{i} + 1}{(z+1)^2}
\end{equation}
with
\begin{equation}
\psi_{0}(t_{i})=\dfrac{24 \pi}{5}G\mu v_{s}\gamma_{s}(z(t_{i})+1)^{-0.5}t_{0}
\end{equation}
and with $z_i$ and $z$ redshifts corresponding to the times $t_{i}$ and $t$, respectively.

This result is in agreement with what is expected from the linear theory of cosmological perturbations. The mass per unit comoving area grows linearly with the scale factor although in this case the wake is described in an idealized situation of matter segments.

Concerning the different signatures of the cosmic string wakes, they correspond to extra-polarization rectangles in the case of the CMB photons or temperature brightness variation rectangles in the case of the 21cm line in the sky of dimension: $c_{1}t_{i}\times t_{i}\gamma_{s} v_{s}$.

\section{Cosmic String Wake signature in CMB polarization} \label{sec3}

In this section, we present a brief description of the polarisation signal from cosmic string wakes, the full derivation of which can be found in \cite{Holder1}. The purpose of this section is to draw a link between the matter inside the wake which contains free electrons and the polarization signal produced by CMB photons interacting with these electrons.

A cosmic string wake is a region of over-density of matter that contains free electrons. The photons coming from the recombination time acquire an extra-polarization when they pass in the wake. This polarization amplitude depends on the height of the wake, the density of free electrons inside the wake when the photons pass through but also on the Thomson cross section $\sigma_{T}$. The magnitude of polarization from the CMB radiation passing through a wake at time $t$, the wake having been created at time $t_{i}$ is: 
\begin{equation}\label{eq:polarization_signal}
P\simeq f\rho_{c}(t_{0})\frac{\Omega_{B}}{m_{p}}\dfrac{(z(t)+1)^{2}}{(z(t_{i})+1)^{-0.5}}(t_i)\sigma_{T} G \mu v_{s} \gamma_{s} t_0 Q
\end{equation}
where $Q$ is the temperature quadrupole,  $m_P$ the proton mass, $f$ the ionization fraction at the time when the photons cross the wake, and  $\Omega_{B}$ the baryon fraction. The constant ${\cal{C}}$ is
\begin{equation}
{\cal{C}} \, = \, \dfrac{24\pi}{25}\left(\dfrac{3}{4\pi}\right)^{1/2} \, .
\end{equation}
Also, we have expressed the baryon density  and the length scale of the wake at the time $t$ (which are relevant for the scattering) in terms of the present critical energy density $\rho_{c}(t_0)$ for a spatially flat universe and the present time $t_0$. Note that the polarization signal is not uniform across the wake. It vanishes at the line in the sky corresponding to the tip of the wake and increases linearly as we go away from the tip. The above value is the average value. 

In order to get an order of magnitude estimate we can insert the values of $\sigma_T$, the proton mass, the critical energy density $\rho_c(t_0)$ and $t_0$. Setting $\Omega_B = 10^{-1}$ and $f = 10^{-4}$ \cite{Holder3} we obtain
\begin{equation}
\frac{P}{Q} \, \sim \, (G\mu)_7 v_{s}\gamma_{s} (z(t)+1)^{2} (z(t_{i})+1)^{1/2} 10^{-13} \, ,
\end{equation}
where $(G\mu)_7$ is the value of $G\mu$ in units of $10^{-7}$. Note that the signal increases quadratically as a function of the redshift when the photons cross the wake, while the dependence on the wake formation time is weaker. 

The signal of a string oriented along the x-axis and moving downwards along the y-axis is shown in Figure 2. Note the linear increase in the signal in y-direction. The maximal signal is about $3 \times 10^{-8} {\rm mK}$. The signal is for a wake with $v_s \gamma_s = 1$ and $G\mu = 3 \times 10^{-7}$ created at the time of recombination (this determines the angular extent of the signal) for which the CMB photons travelling to us are crossing the wake at a redshift of $z = 20$. This redshift was chosen since it is in the range of 21-cm experiments which are in construction.

\begin{figure}[h]
    \includegraphics[width=8cm]{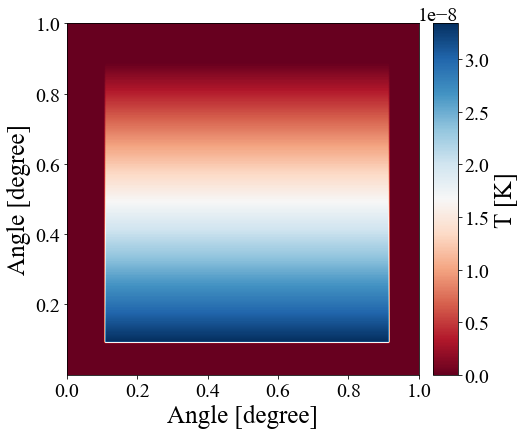} 
    \caption{Temperature in K of the B-mode polarization signal of a cosmic string wake with string tension $G\mu=3\times 10^{-7}$. This wake was created at $z_i = 1100$ and the CMB photons were polarized by this wake at $z = 20$. This wake occupies a patch of sky measuring 1 square degree.}
\label{fig:pol_wake_only}
\end{figure}

\section{Cosmic String Wake signature in 21 cm Brightness temperature} \label{sec4}

In this section we outline the signal in the brightness temperature of the 21-cm line produced by cosmic string wakes. Here we provide only a short review of the physics at play, and for readers interested in a more detailed discussion on the 21-cm signal of cosmic string wakes we refer to \cite{Holder1} and \cite{Maibach}.

As described above, cosmic string wakes are matter overdensities which form behind moving long string segments. Before the time of reionization, the baryonic matter in these overdensities consists mainly of neutral hydrogen. If the gas temperature of this neutral hydrogen is lower than the temperature of the CMB photons (which it will be at higher redshifts and for the low values of the string tension which we are interested in), then CMB photons passing through the wake will be partially absorbed, exciting the 21-cm hyperfine transition of neutral hydrogen. This will lead to an absorption feature in the 21-cm sky: the 21-cm temperature will be lower in regions of the sky where the CMB photons have passed through a string wake. 

The 21-cm signal of a cosmic string wake corresponds to a ``thin slice of a pie'' in the three-dimensional 21-cm sky. The angular extent of the feature will be given by the comoving Hubble radius when the string segment forms, and the thickness of the feature in redshift direction is determined by $G\mu$ (and be the abovementioned comoving Hubble radius). At a fixed redshift, the string wake signal appears as a rectangle in the sky with the same angular size and same position in the sky as the B-mode polarization signal. As in the case of the B-mode polarization signal, there is a linear gradient in the intensity of the 21-cm signal, with the amplitude along the line in the sky corresponding to the tip of the wake being lowest (see Figure 3).

The amplitude and redshift extent of the string wake signal depends on the temperature and density of the gas around and inside the wake (see \cite{Oscar2} for a detailed discussion). If the temperature of the gas $T_g$ in the surrounding space is low, then the temperature of the matter falling into the wake will be determined by shock heating, will acquire a value $T_K$ which can be calculated from virialization, and the matter will remain confined to the wake. This will be a good description provided $T_K > 3 T_g$. In the other case matter also heats the wake but the wake will be more diffuse, i.e. the wake thickness will be greater but its density lower. Note that the projected 21-cm signal will be the same. 

The local intensity of the 21-cm absorption signal for photons passing through a wake formed at the time of recombination is \cite{Holder2, Oscar2, Maibach}
\begin{equation}\label{eq:21cm_signal}
\delta T_{b}(\nu)=[17\rm{mK}]\dfrac{x_{c}}{1+x_{c}}(1-\dfrac{T_{\gamma}}{T_{K/g}})      \frac{(1+z)^{0.5}}{2\sin^2(\theta)}\frac{n_{HI}^{wake}}{n_{HI}^{bg}} \, ,
\end{equation}
where $x_c$ is the collision coefficient, $z(t)$ is the redshift when the photons pass through the wake, and $T_{\gamma}$ is the temperature of the CMB photons at that time. $\theta$ describes the angle between the normal to the wake and the directionn the photons are travelling on their way to us. Since we are interested in the order of magnitude of the effects we will simply use $2\sin^2(\theta) = 1$. The above formula also contains the number densities $n_{HI}$ of neutral hydrogen. Note that 
\begin{equation}
    T_{K/g}= {\begin{cases}
  T_K  &  T_{K}>3T_{g}\\
 3T_{g} &  T_{K}\leq 3T_{g}
\end{cases}}
\end{equation}
and
\begin{equation}
    \frac{n_{HI}^{wake}}{n_{HI}^{bg}} = {\begin{cases}
  4  &  T_{K}>3T_{g}\\
 1+\frac{T_K}{T_g} &  T_{K}\leq 3T_{g}
\end{cases}}
\end{equation}

Note that the brightness amplitude at a local point in the 21-cm sky does not depend on $G\mu$. The extent of the signal in redshift direction, however, grows linearly with $G\mu$, and hence a signal integrated over a finite redshift range will depend on the string tension. Note also the the local amplitude of the signal is large compared to the amplitude of features produced by primordial Gaussian fluctuations. This is in marked contrast to the amplitude of the string-induced polarization signal which is must smaller than the signal from Gaussian noise. 

\begin{figure}[h]
    \includegraphics[width=8cm]{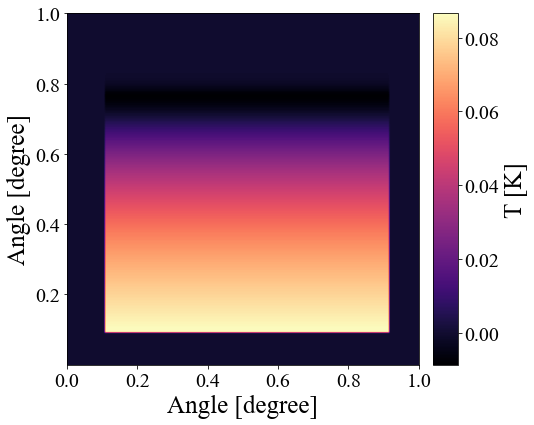} 
    \caption{21cm brightness temperature of a cosmic string wake with string tension $G\mu=3\times 10^{-7}$. This wake was created at $z_i = 1100$ and the 21cm photons were emitted from this wake at $z = 20$. This wake occupies a patch of sky measuring 1 square degree.}
    \label{S21_3_0_e7.png}
\end{figure}

\section{Methods}\label{sec5}

This section is the heart of the article. Here we explain the methods that we use to simulate our cosmic string wake populated fields, as well as the methods used to then extract those non-Gaussian signals from the sky maps. The first statistical tool to come to mind when dealing with a non-Gaussian signals are higher order correlation functions such as the three-point correlation function or its Fourier transform, the bispectrum. While for many scenarios this is an appropriate tool, the special geometry of wakes in combination with projection effects make it difficult to choose a universal shape for computing three-point statistics. Instead, we make use of position space cross-correlations in addition to matched filtering to explore whether one can detect cosmic string wakes during the cosmic dark ages. 

\subsection{Mock Observation and Analysis}\label{sec:5A}

We first produce mock observations consisting of Gaussian realizations of the CMB B-mode polarization and 21-cm fields at $z = 20$ of a Planck-normalized LCDM cosmological model \cite{Planck} onto which we add the wake signal. To produce the Gaussian backgrounds we sample the power spectra of these probes in Fourier space and inverse Fourier transform them to produce position space maps. For the CMB B-mode polarization, we sample its primordial spectrum provided by the \texttt{CAMB} code \cite{camb}. On the 21-cm side, we sample the matter power spectrum also provided by \texttt{CAMB} which is consistent with the expected dark ages signal. We re-scale the multipoles in order to obtain a spectrum in wave number modes from which we sample.  

For this work, we generate two different sets of mock samples, one without cosmic string wakes and one with cosmic string wakes, in order to be able to evaluate the detectability of cosmic string wakes against the null result. The no-signal sample is simply the mock maps as described in the previous paragraph. For the samples with cosmic string wakes we compute the wake signal for wakes produced at $z  = 1100 $ in B-mode polarization and in 21-cm  from equations \ref{eq:polarization_signal} and \ref{eq:21cm_signal} respectively, and we add the maps to the no-signal maps. An example of a resulting B-mode polarization map is shown in the middle panel of the top row of Figure 4 and a resulting 21cm map is shown in the middle panel of the bottom row of Figure 4. We are taking the redshift at which the photons cross the wake to be $z = 20$. As can be easily seen, the wake signal is dominant to the 21 cm background at this redshifts while the wake signal is by many orders of magnitude sub-dominant to the B-mode polarization background.

\begin{figure*}
\label{6_maps}
    \includegraphics[width=7.3in]{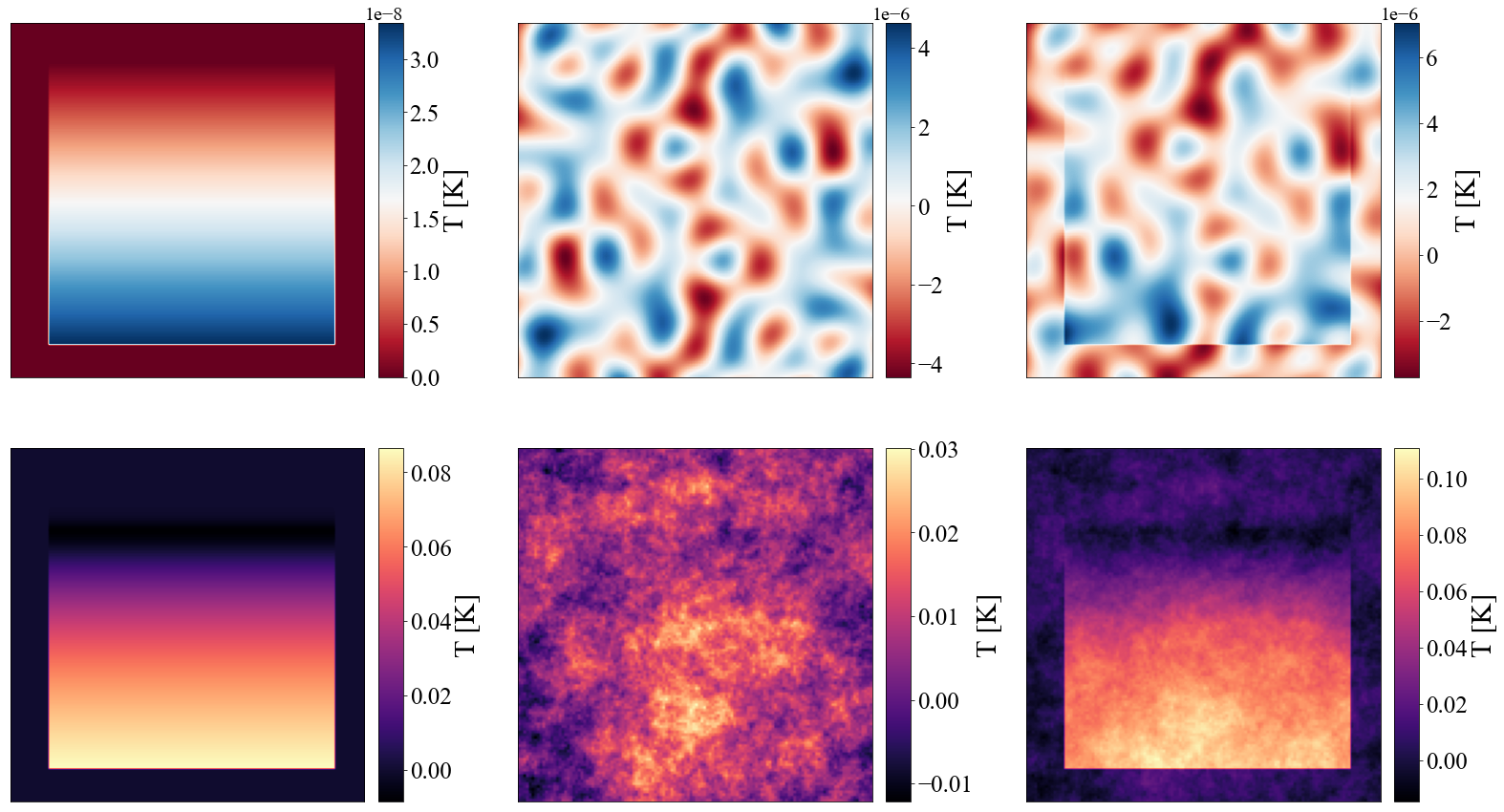} 
    \caption{Shown here are all of the simulation components for both the cosmic string wakes with string tension $G\mu=3\times 10^{-7}$. Once again, these wakes were created at $z_i = 1100$ and emission occurred at $z = 20$. This wake occupies a patch of sky measuring 1 square degree. The top row of maps depicts the CMB related simulation components. In the top left map, the cosmic string wake signal from B-mode polarization in K is shown. The middle map of the top row is the simulated B-mode polarization background where Gaussian realizations were drawn from the B-mode power spectrum from \texttt{CAMB}. The top right panel shows the full simulation with the cosmic string wake superimposed on the CMB. It should be noted that in order to make the wake signal visible to the eye, the wake signal was boosted by a factor of 100. This should also exemplify just how dim the B-mode signal is in comparison to the background. The bottom row shows corresponding simulation components for the 21cm observations. The bottom left map is the 21cm signal from cosmic string wakes in K. The middle map in the bottom row is the 21cm background where Gaussian realization were drawn from the matter power spectrum from \texttt{CAMB}. The bottom right panel shows the full simulation with the cosmic string wake superimposed on the 21cm background.}
\end{figure*}

According to the string scaling solution, the most numerous wakes are those produced at recombination (since the comoving Hubble radius is the smallest and there are a fixed number of wakes produced at each time step per comoving Hubble volume at the formation time). The angular extent of the induced wake signals in the sky is of the same order, i.e. of the order of $!$ degree. We will hence focus on 1 square degree Gaussian noise sky maps onto which we place the signal of string wakes. In total, we produce 10,000 realization with and without string wake signatures. This corresponds to a total sky coverage of $100 \times 100$ square degrees. We consider this to be realistic sky coverage for future experiments.  

Once these mock catalogs are produced, we proceed to perform our cross correlation analysis where we compute the cross-correlation coefficients between the B-mode maps and the 21-cm brightness temperature maps with and without wakes. We then average together the cross-correlation coefficients computed from all our realizations in order to obtain an average statistic. We found that it is not possible to extract the string signal for values of $G\mu$ lower than the current upper bound by looking at the individual cross-correlation coefficients,  even with full sky coverage. We therefore proceeded to apply matched filtering to the average cross-correlation coefficient map in order to attempt to extract the string signal. In the following subsections, we describe our analysis tools in more detail.

\subsection{2D cross-correlations}\label{sec:5B}

We use cross-correlations to tease out the cosmic string signal in the simulated maps. Cross-correlations allow one to pick out similarities between two functions $f$ and $g$. In the case were $f$ and $g$ are both bivariate functions, the cross-correlations coefficients are defined as follows:
\begin{equation}\label{eq:cij_function}
C(i,j)=\int \int f(x,y)g(x+i,y+j) \, \mathrm{d}x\mathrm{d}y
\end{equation}
These cross-correlations are widely used for image recognition and template matching. By convolving and image with a template, the cross-correlation coefficients, $C(i,j)$, provide information as to where the image is most similar to the template. 

In the case of two dimensional images $F$ and $G$ each in the form of an $n\times n$ matrix of pixels, the formula equation \ref{eq:cij_function} is discretized in the following way, 
\begin{equation}\label{eq:cij_discrete}
C_{ij}=\sum_{x,y=-\frac{n}{2}}^{n/2} (F_{xy}-\langle F \rangle)(G_{(x+i)(y+j)}-\langle G \rangle)
\end{equation}
where $\langle F \rangle$ denotes mean value of $F$ and $\langle G\rangle$, the mean value of $G$.

\begin{figure*}
\label{C(i,j)_3_0e7.png}
    \includegraphics[width=18cm]{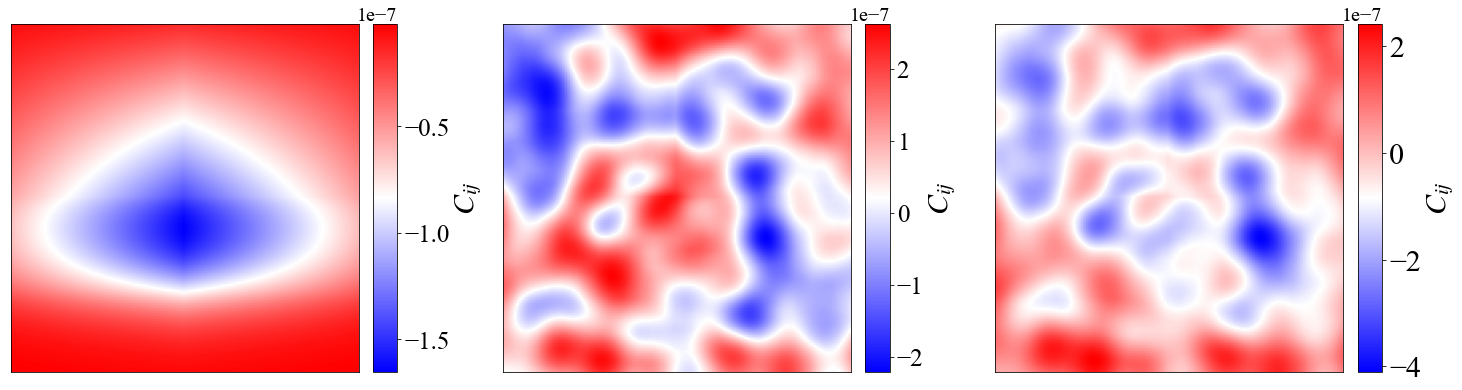} 
    \caption{Cross-correlation coefficient matrix for 21cm and B-mode observations. The left panel shows the cross-correlation matrix for the naked 21cm and B-mode wake signals. A clear anti-correlated signal is present in the cross-correlation due to the correlation between the B-mode emission and the 21cm absorption in the cosmic string wake. The middle panel shows the cross-correlation matrix for maps without wakes present. As expected, there is no signal. In the right panel, the cross-correlation matrix for maps with both a Gaussian background and a string wake signal is shown. Again, no discernable signal is present. The wakes simulated here have a string tension of $G\mu=7\times 10^{-8}$. These wakes were created at $z_i = 1100$ and emission occurred at $z = 20$. The wakes occupy a patch of sky measuring 1 square degree.}

\end{figure*}

By construction, the mean value of the Gaussian fluctuations of the polarization and the 21-cm maps are zero. The cosmic string wake signal, while constituting an overdensity, also has surrounding underdensities as a result of having matter funnelled into the wake (this also insures agreement with the Traschen integral constraints on cosmological perturbations \cite{Traschen}). Therefore, relative to the background, the entire cosmic string wake signal, including the surrounding underdensities, also has vanishing mean. In our analysis, we therefore do not explicitly subtract any mean in computing the sums $C_{ij}$.

Note that we do not include any rescaling (normalization) of the cross-correlatdion coefficients). This is justified since we are comparing maps with similar overall strength, given that the wake signal (which is the difference between the two maps) is a sub-dominant contribution in the B-mode polarization map.
 
For two-dimensional maps, cross-correlation coefficients allow to identify similarities in the values of the pixels of an image. If we have two similar images positioned at different locations of the maps, then the $(i, j)$ values for which the cross-correlation coefficients are maximal give us the shift in the images in the $(x, y)$ directions. In our case, the wake images are located at the same location in the sky, and hence the contribution of the wake to the cross-correlation map will be peaked at $(i, j) = (0, 0)$. The length in the $(i, j)$ plane over which the cross-correlation signal decays is related to the extent of the image.  

The left column of Figure 5 shows the cross-correlation matrix between CMB B-mode polarization and 21-cm signal of the pure cosmic string wake signals. Since the wake features are at the same position in the sky, the peak of the cross-correlation coefficients is in the center of the map. Since the features are rectangles in the sky with axes chosen to coincide with the coordinate axes, the images show spikes extending along the coordinate axes. For a cosmic string wake with a general orientation, the spikes would not be along the coordinate axes. Since the amplitude of the string-induced B-mode polarization signal is very low, the amplitude of the cross-correlation coefficients is also small.

However, if we consider maps containing the Gaussian noise plus the wake signal, we find that, for values of the string tension smaller than the current upper bound, and for a redshift $z = 20$ when our past light cone crosses the wake, the wake signal is not discernible by eye, even after stacking all of the $10,000$ maps which we generate (the string wake cross-correlation maps are identical modulo orientation, but the noise maps vary with the realization of the Gaussian noise). We see this in the middle and right panels of Figure 5. The middle panel shows the cross-correlation coefficients of maps without wakes while the rightmost panel shows the cross-correlation coefficients for maps with wakes (both averaged over the realizations of the random background noise).

\subsection{Matched filter}\label{sec:5C}

\begin{figure*}
  \includegraphics[width=0.95\textwidth]{{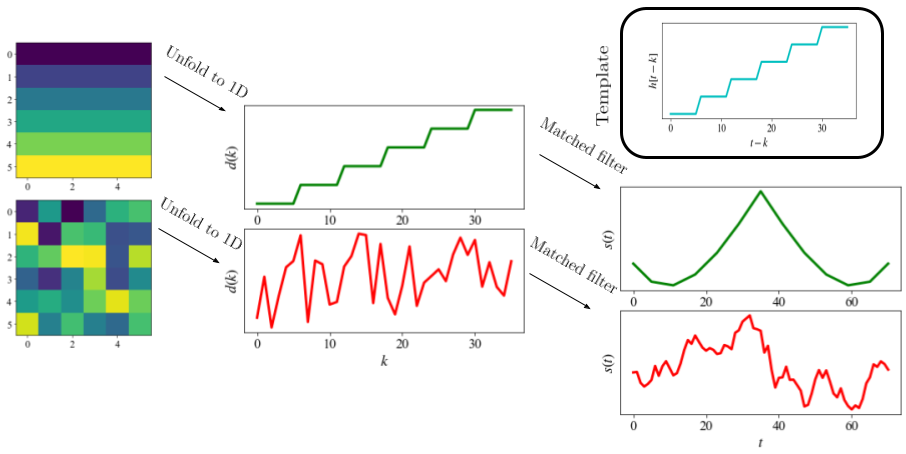}}
  \caption{This flowchart depicts the process of taking  the cross-correlation matrix, unfolding it to a 1D array, and applying a matched filter. The top row, from left to right, shows the process of unfolding an increasing signal and applying a matched filter where the template is itself that same increasing signal. We see in the top $s(t)$ curve that as the template slides over the data $d(k)$, there is a strong match signal when perfect overlap is achieved. The bottom row shows the same unfolding and mathed filtering process but for a matrix made up of Gaussian random noise. It is easy to see that in the bottom curve labelled $s(t)$, nowhere was the template signal found in the data.}
\end{figure*}

As discussed above, the cross-correlation coefficients alone are insufficient for extracting the wake signal with a reasonable number of maps. We proceed to explore the use of matched filtering on these average $C_{ij}$ matrices. Matched filters are widely used to detect the presence of a signal or template in a data set. 

We will first turn the $n \times n$ cross-correlation matrix into a one-dimensional data stream of length $n^2$ by unfolding the rows from top to bottom into a vector. The matched filter correlation vector between the data and the predicted template can be defined as
\begin{equation}
    s(t) = \sum _{k = -\frac{n^2}{2}}^{\frac{n^2}{2}} h[t-k]\times d[k] 
\end{equation}
where $h$ is the template and $d$ is the data from which one is trying to extract the signal. 

The templates of correlations, $C_{ij}$, which we are trying to extract are obtained from the $C_{ij}$ for maps containing only the wake signals. The one-dimensional templates are then obtained by unfolding the $C_{ij}$ matrix into a one-dimensional array. We then apply this filter to the averaged $C_{ij}$ matrix that has been unfolded in the same manner as the template. We apply this filter to the cross-correlations computed from maps with both a wake and a background contribution as well as maps that contain only the Gaussian realized fluctuations. We expect to be able to recover the signal from those maps containing wakes but, of course, to get a null result when using the template on maps without wakes. What follows are the results of using the analysis methods described here, allied to the mock catalogs which we have generated.

\section{Results}\label{sec6}

Here we focus on cosmic strings with a tension of $G\mu = 7\times 10^{-8}$, slightly lower than the current robust upper bound. If we assume that the cosmic string loop distribution scales and that gravitational radiation is the main energy loss mechanism for a string loop, then a significantly tighter upper bound can be derived from millisecond pulsar timing experiments \cite{MSP}. Such a bound, however, is not robust since it depends on unproven assumptions about the string loop distribution, assumptions which are in conflict with the results of some field theory simulations \cite{Hind}. We first compute the average cross-correlation coefficients for simulations with and without cosmic string wakes. These cross-correlation matrices are then unfoled into 1D arrays for matched filtering.  As is evident, the string wake signal is invisible by eye before matched filtering.

With matched filtering, however, the string signal can be cleanly extracted, as is shown in Figures 7. The top panel of Figure 7 shows the template used in the matched filtering process. This template is simply the unfolded $Cij$ matrix computed from maps with only the 21cm and B-mode polarization signals present (left panel of Figure 5 unfolded into a 1D array). The bottom panel of Figure 7 shows the matched filter results as the function $s(t)$. The red curve shows the results for the background noise, averaged over the realizations of the noise. The spread of the data points provides a good measure of the statistical error. The curve is not symmetric and is clearly noisy. The amplitude of the signal at a given position of the vector can vary from realization to realization. The yellow curve in the bottom panel of Figure 7 depicts the same results for the pure string wake signal, and the green curve in the bottom panel of Figure 7 is the result for maps containing both background and cosmic string signal (again averaged over realizations of the noise. The string wake signal in the matched filtering function is symmetric, has a positive sign, and is larger in amplitude than the signal for the noise maps. Most importantly, the results for the noise maps have a very different shape than what is obtained for pure string wake maps. When the matched filtering algorithm is applied to the maps containing both background and string signal, the pure string signal is clearly visible.

\begin{figure}[h!]
\includegraphics[width=9.2cm]{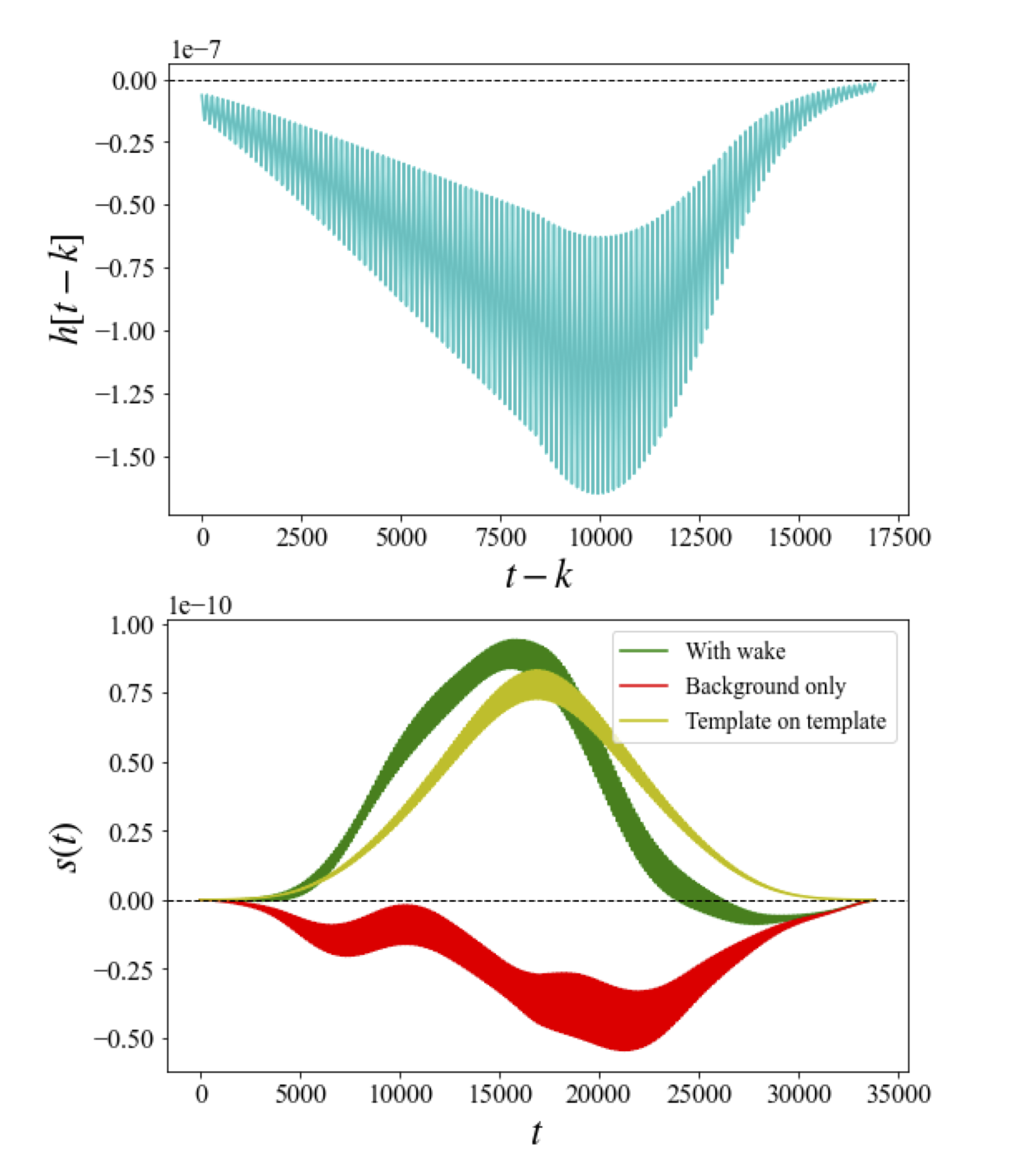} 
\caption{The top panel shows the template used in the matched filtering process. This template is simply the unfolded $Cij$ matrix computed from maps with only the 21cm and B-mode polarization signals present with a string tension of $G\mu = 7\times10^{-8}$ (left panel of Figure 5 unfolded). The bottom panel shows the matched filtering results when this template is applied to 3 different sets of maps. The yellow curve shows the matched filtering signal when the template is convolved with itself. The green curve shows the matched filtering result when the template is applied to maps with both the Gaussian background and the wake signal present. The red curve is the matched filter results for the template applied to maps where only the Gaussian backgrounds are present. As expected the template is found in the maps with cosmic string wakes present, and it is not found in maps without wakes present. }
\end{figure}





We have thus established that, using matched filtering, the signal of string wakes can be extracted from the ``background noise'', i.e. the signals from the Planck-normalized LCDM model, at a statistically very significant level for values of the string tension slightly lower than the current robust upper bounds. It is clear from Figure 7 that the string signal can also be extracted for values of $G\mu$ significantly lower than the value we have used. However, our analysis involves a number of idealizations which should be removed before we can make definite statements about the level of $G\mu$ which in fact can be reached with our analysis. For example, we have taken the string wake to have a fixed orientation over all realizations of the background noise. Different wakes, however, will have different orientations, and this should be taken into account.

\section{Discussion and Conclusion}

We have shown that the signals of a cosmic string wake can be extracted from background cosmology noise by considering  cross-correlation maps between CMB B-mode polarization and 21-cm temperature anisotropy maps, the latter considered as a two-dimensional map at a fixed redshift. Specifically, we have shown that the signals from a string wake with tension $G\mu = 7 \times 10^{-8}$, i.e. lower than the current robust upper bounds, can be extracted for string wakes generated at the time of recombination which are crossed by our past light cone at a redshift of $z = 20$. Wakes produced at recombination are more numerous than those produced later, which is a good motivation for considering these wakes. The reason for choosing wakes which intersect our past light cone at $z = 20$ is that this redshift is higher than the redshift of reionization when nonlinear effects will start to disrupt the wake signal \cite{Disrael2}, and that it is in the range of some current and planned experiments such as the MWA \cite{MWA} and the SKA \cite{SKA}.

B-mode polarization and 21-cm temperature maps are each dominated by foregrounds. One of the motivations for considering our cross-correlation analysis is that the signal shapes of the wakes are highly correlated in the two maps, but the contributions of the noise are not. Hence, ideally one could hope that the noise effects would cancel out upon spatial averaging, while the signal will remain.

The main problem of the analysis is that the string-induced contribution to B-mode polarization from wakes which are past light cone crosses at low redshifts is minuscule compared to the signal of the usual LCDM fluctuations. This leads to the fact that the string wake signal (for the parameters we have chosen) is not visible in the individual cross-correlation maps. However, we have shown that the signal can be extracted using matched filtering.

If we were to consider wakes which our past light cone crosses at much earlier times, the B-mode signal would be much stronger since it scales as $(z + 1)^2$. The 21-cm signal is also an increasing function of $z$. Hence, in the absence of foreground noise, much stronger constraints on $G\mu$ would be achievable. However, since the noise in 21-cm maps increases rapidly as a function of redshift, we would need to be able to subtract such noise to high precision. Note that there are observational 21-cm projects that plan to reach redshifts close to that of recombination.

\section{Acknowledgements}

This research is supported in part by funds from NSERC and from the Canada Research Chair program as well as the Fonds de recherche du Québec – Nature et technologies (FRQNT). We wish to thank Adrian Liu for discussions.


\end{document}